\documentclass[preprint,12pt]{revtex4-1}
\usepackage{multirow}
\usepackage[table]{xcolor}
\usepackage{rotating}
\usepackage{amsmath}
\usepackage{bm}
\usepackage{array}
\usepackage{graphicx}
\usepackage{dcolumn}   
\usepackage{amssymb}
\usepackage[english]{babel}

\begin{document}
\title{Analytical solutions to slender-ribbon theory}
\date{\today}
\author{Lyndon Koens\footnote{lmk42@cam.ac.uk}, and Eric Lauga\footnote{e.lauga@damtp.cam.ac.uk}}
\affiliation{ Department of Applied Mathematics and Theoretical Physics, University of Cambridge, Wilberforce Road, Cambridge CB3 0WA, United Kingdom}

\begin{abstract} 
The low-Reynolds number hydrodynamics of slender ribbons is accurately captured by slender-ribbon theory, an asymptotic solution to the Stokes equation which assumes that  the three  length scales characterising  the ribbons are well separated. We show in this paper  that  the force distribution across the width of an  isolated ribbon located in a infinite fluid  can be determined analytically, irrespective of the ribbon's shape.  This, in turn, reduces the surface integrals in the slender-ribbon theory equations to a line integral analogous to the one arising in  slender-body theory to determine the dynamics of filaments.  This result is   then used to derive analytical solutions to the motion of a rigid plate ellipsoid and a ribbon torus and to propose a  ribbon resistive-force theory,  thereby extending  the resistive-force theory for slender filaments.

\end{abstract}
\maketitle
\def\v{\vspace{2cm}}

\section{Introduction}
Stokes flows problems are appealing to mathematicians because of the large array of asymptotic tools available to solve them  \cite{Kim2005}. There are, however, relatively few exact  solutions \cite{Brady1988,Kim2005,Leal2007}. These analytical solutions are usually found in one of three ways: (a) solving the Stokes equations  directly \cite{Kim2005, Happel1981}; (b)  analytically inverting a boundary integral formulation \cite{LAMB1932}; or (c) using judiciously-placed flow singularities \cite{Chwang2006}.  
Both methods (a) and (b) require the use of a clever coordinate system that matches the geometry of the problem (such as spherical,  ellipsoidal, toroidal or bi-spherical coordinates), in order to derive analytical solutions; in contrast, method (c), a singularity representation, requires only a guess at the type of flow singularities needed and where these singularities are located. 

The boundary integral formulation (b) is very powerful and is often used for numerical calculations \cite{Pozrikidis1992}, while the singularity method (c) lends itself better to series expansions or situations where a numerical discretisation of the body surface would become difficult  \cite{Johnson1979a, Batchelor2006}. 
An example of such a shape, for which discretisation is difficult, is a long thin cylindrical filament since an appropriate a computational mesh needs to resolve both the width and the length of the filament. Slender filaments abound in the biological world, for example the flagella that many microorganisms use to propel themselves \cite{Lauga2009}. Therefore it is important to have appropriate models to capture their low-Reynolds number dynamics.  

The main mathematical technique used to accurately capture the hydrodynamics of slender filaments in a flow is called slender-body theory (SBT)\cite{1976, Johnson1979, Sol1976}.  This technique relies on slender-body having two regions of behaviour: a local cylindrical region that scales with the filament's width, $2 r_{b}$, and a long range hydrodynamic interaction region that scales with the filament length, $2\ell$,. These two regions are then matched together to capture the total flow. This matching can be done in a number of ways, thereby creating multiple versions of the theory. For example Keller and Rubinow's SBT \cite{Sol1976} matches the Stokes flow around an infinite cylinder, method (a) above, to a line of stokeslets (point forces), method (c). This creates a physically intuitive version of SBT that is accurate to $r_{b}/\ell$. Alternatively Johnson's SBT \cite{Johnson1979} mathematically represents the total flow around a slender filament by placing a series of singularity solutions to the Stokes equations along the filament's centreline, method (c), and then expanded the solution in orders of the thickness over length.   Though less intuitive, this method also determines the structure of the higher order corrections exactly.  This enabled Johnson to show that his leading order equation predicted the force accurately to order $(r_{b}/\ell)^{2} \ln(r_{b}/\ell)$. Hence Johnson's SBT is considered the most accurate. Specifically he found that the leading-order velocity of the filament at arclength $s$ along the centerline,  $\mathbf{U}(s)$, is given by 
\begin{eqnarray} 
8 \pi \mu \mathbf{U}(s) &=&  \int_{-\ell}^{\ell} \left[ \frac{\mathbf{I}+ \mathbf{\hat{R}}_{0} \mathbf{\hat{R}}_{0}}{|\mathbf{R}_{0}|}\cdot \mathbf{f}(s') -\frac{\mathbf{I}+ \mathbf{\hat{t}\hat{t}}}{|s'-s|} \cdot \mathbf{f}(s) \right] \,d s' \notag \\ 
&&+\ln\left(\frac{4 \ell^{2}(1-s^{2})}{r_{b}^{2} \rho(s)^{2} e}\right)\left(\mathbf{I}+ \mathbf{\hat{t}  \hat{t}}\right) \cdot \mathbf{f}(s)+2\left(\mathbf{I}- \mathbf{\hat{t}  \hat{t}}\right) \cdot \mathbf{f}(s),\label{SBT}
\end{eqnarray}
where $\mathbf{f}(s)$ is the (unknown) force distribution along the body's centreline  \cite{Johnson1979,Typo}. In the above equation, $e$ is the exponential, $\rho(s)$ is the dimensionless radial surface distribution (so that the surface of the body is located at $r = r_b \rho(s)$),  $\mathbf{R}_{0} = \mathbf{r}(s)-\mathbf{r}(s')$ is the vector between points at $s$ and $s'$ on the centreline and $\mathbf{\hat{t}}$ is the unit tangent to the centreline at location $s$.  Johnson's SBT has been very successful in capturing the hydrodynamics of slender filaments in a variety of settings   \cite{Lauga2009, Koens2014, Tornberg2006, Tornberg2004, Yariv2013, Barta1988} and can be used to determine the  hydrodynamics of thin prolate ellipsoids \cite{Johnson1979,Tornberg2006} and slender tori \cite{Johnson1979a} analytically. 
The use of slender-body theory, combined with accurate experimental measurements, has significantly improved our understanding of the motion of swimming microorganisms \cite{Lauga2009, Chattopadhyay2006, Dombrowski2009, Wolgemuth2000}. This understanding has then prompted the scientific community to create artificial microswimmers \cite{Zhang2010,Maggi2015a,Ao2014a}. 

As a difference with biological swimming cells, many artificial swimmers use slender appendages in the shape of ribbons rather than filaments \cite{Zhang2010,Kohno2015}.  These  slender-ribbons are seen to exhibit different physics to a slender-filament \cite{Pham2015, Xu2015a, Dias2014, Keaveny2011}, thereby requiring new tools to mathematically model their behaviour. Fundamentally, many problems in the natural or industrial world are concerned with slender bodies shaped like ribbons, including swimming sheets \cite{Diller2014, Montenegro-Johnson2016}, curling ribbon membranes \cite{Arriagada2014, Tadrist2012} and carbon nano-ribbons  \cite{Bandura2016}. 
Recently we derived a slender-body theory-like expansion to describe the hydrodynamics of ribbons \cite{Koens2016}. This theory, which we called slender-ribbon theory (SRT), was seen to give accurate numerical results and capture the dynamics of ribbon shaped artificial microswimmers. While the derivation of the method was all done analytically, the final result was a double integral equation which had to be inverted numerically. Since, in the case of filaments,  some analytical solutions to SBT exist, we consider in this paper the extension to the case of ribbons and show that analytical solutions to SRT do exist as well.

Specifically we show in this paper that the force distribution across a slender ribbon's width can be solved exactly for any arbitrary isolated ribbon. This significantly simplifies the general SRT equations, reducing the surface integrals to a line integral. By considering the hydrodynamics and settling behaviour of a long flat ellipsoid and a ribbon torus we show that the line integrals can be solved exactly in these cases, thus providing analytical solutions.

The paper is organised as follows. In Sec.~\ref{SRT} we briefly summarise the derivation of slender-ribbon theory before discussing the challenges in solving the final integral equations analytically in Sec.~\ref{diff}. We then solve for the force distribution across the ribbons width arbitrarily (Sec.~\ref{fdis}) and use this result to simplify the general SRT equations for an arbitrary isolated ribbon (Sec.~\ref{rSRT}). Finally in Sec.~\ref{analytic} we analytically determine the rigid-body hydrodynamics and settling behaviour of a long flat ellipsoid (Sec.~\ref{arod}) and of a ribbon torus (Sec.~\ref{atorus}).

\section{Slender-ribbon theory} \label{SRT}
\subsection{Finite ribbons}
\begin{figure}[t]
\centering
\includegraphics[width=0.8\textwidth]{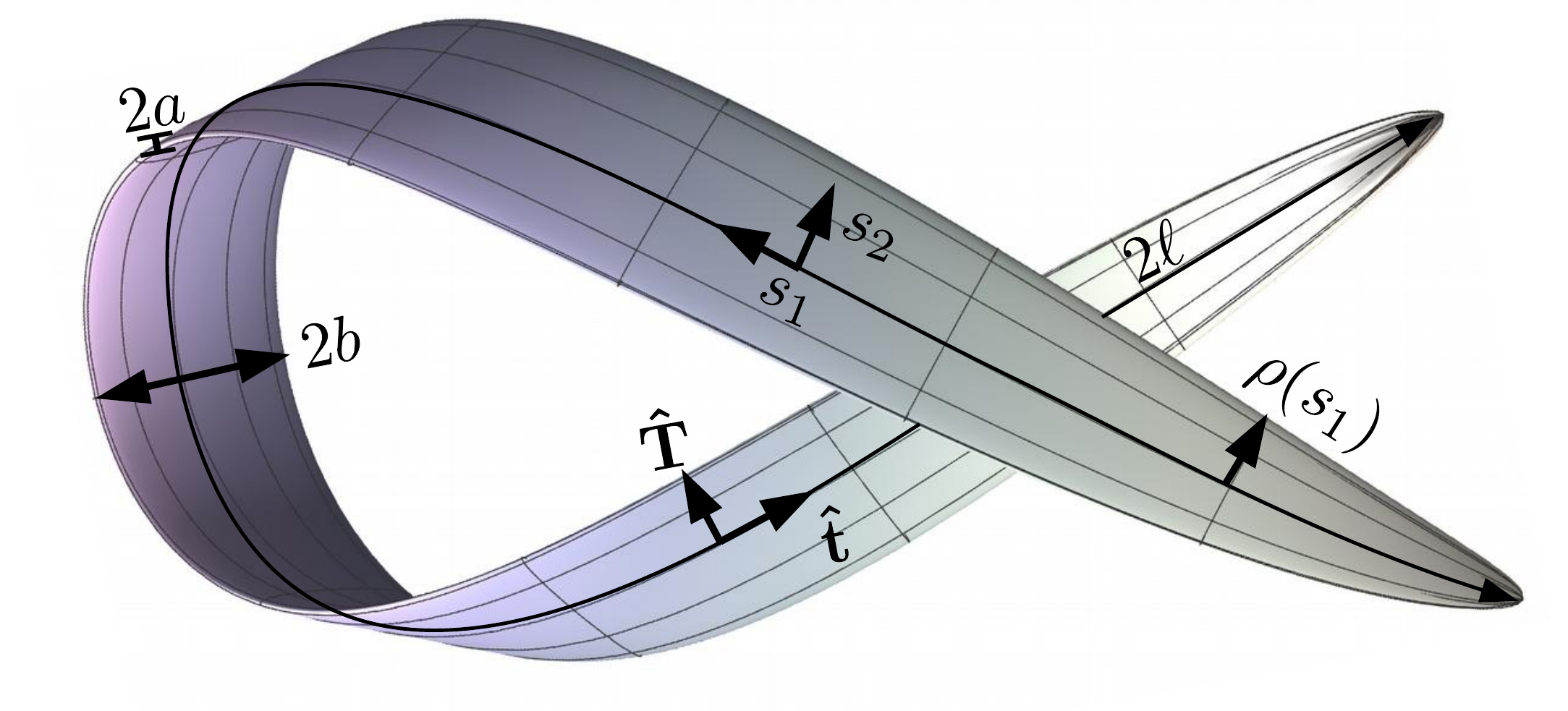}
\caption{Sketch of a slender-ribbon of length $2\ell$, width $2b$ and thickness $2a$:  $\mathbf{\hat{t}}$ is the tangent vector to the ribbons centerline, $\mathbf{\hat{T}}$ is a unit vector pointing in the direction of the ribbons width, $\rho(s_{1})$ is the cross-sectional shape of the ribbon width, and  $s_{1}$ and $s_{2}$ are the arclengths along the ribbon's centreline and width, respectively.}
\label{fig:ribbon}
\end{figure}

Consider a  slender ribbon defined by its centreline, $\mathbf{r}(s_{1})$, and a unit vector $\mathbf{\hat{T}}(s_{1})$ which is perpendicular to the centrelines tangent vector, $\mathbf{\hat{t}}(s_{1})$, and points in the direction of the ribbon's width (see sketch in Fig.~\ref{fig:ribbon}). The slenderness of the ribbon is enforced by assuming that the centreline length of the ribbon, $2\ell$, is much larger than the width, $2 b$, which itself is much larger than the thickness $2 a$, i.e.~$\ell\gg b\gg a$.  The hydrodynamics of the ribbon is then determined by placing stokeslet singularities (point forces) over an imaginary plane which lies within the ribbon and expanding the resulting velocity on the ribbon surface in orders of $b_{\ell} \equiv b/\ell\ll 1$ and $a_{\ell}\equiv a/\ell\ll 1$ \cite{Koens2016}.  

This expansion is performed similarly to that of Johnson's slender-body theory \cite{Johnson1979, Gotz2000} in order to accurately quantify the error and higher order corrections of the expansion \cite{Errors}. Similarly to all slender-body theories, slender-ribbon theory also exhibits multiple regions of behaviour. However unlike slender-body theory, three relevant regions are found: an outer region, capturing the long range physics of the fluid, a middle region, where the body is locally a flattened cylinder, and an inner region, where the body is locally an infinite flat sheet. In accounting for each of these regions, the relevant physics from the limits $b_{\ell} \rightarrow 0$ and $a/b \rightarrow 0$ is captured and the result becomes independent of the order of limits taken. Mathematically this is supported by the fact that $a/b$ only occurs in product with $b_{\ell}$ within the expanded functions \cite{Koens2016}. This derivation generates an integral equation, valid to $O(b_\ell)$, with the form
 \begin{eqnarray}\label{6:SREs}
\notag 8 \pi \mathbf{U }(s_{1},s_{2}) &=& \int_{-1}^{1} \,d s'_{1}  \left[ \frac{ \mathbf{I} + \mathbf{\hat{R}_{0}} \mathbf{\hat{R}_{0}}}{|R_{0}|} \cdot \left\langle\mathbf{f}\right\rangle(s'_{1})   -  \frac{ \left(\mathbf{I} + \mathbf{\hat{t}} \mathbf{\hat{t}}\right)}{|s'_{1}-s_{1}|} \cdot \left\langle\mathbf{f}\right\rangle(s_{1})  \right]  \\&&
\notag +  \int_{-1}^{1} \,d s'_{2} \left[\ln\left(\frac{4(1-s_{1}^{2})}{b_{\ell}^{2} \rho(s_{1})^{2} (s_{2}-s'_{2})^{2}}\right)  \left(\mathbf{I} +\mathbf{\hat{t}} \mathbf{\hat{t}} \right) \cdot \mathbf{f}(s_{1},s'_{2})\right]\\
&& +2    \left(\mathbf{\hat{T}} \mathbf{\hat{T}}- \mathbf{\hat{t}} \mathbf{\hat{t}}\right) \cdot \left\langle\mathbf{f}\right\rangle(s_{1}),
\end{eqnarray} 
where $\mathbf{U}(s_{1},s_{2})$ is the velocity on the surface of the ribbon at arclengths ($s_{1}$, $s_{2}$), $\rho(s_{1})$ is the cross-sectional shape of the ribbon width, $\mathbf{f}(s_{1},s_{2})$ is the force distribution over the stokeslet plane,  $s_{1}$ is the arclength along the centreline, $s_{2}$ is the arc-length along the ribbons width, $\mathbf{R}_{0} = \mathbf{r}(s_{1})-\mathbf{r}(s'_{1})$ and  $\left\langle\cdot\right\rangle \equiv \int_{-1}^{1} \,ds_{2}$ denotes the total across the width of the ribbon.  The beyond $O(b_{\ell})$  corrections to this equation are of $O(b_{\ell}^{2})$ or $O(a_{\ell})$ depending on the relative dimensions of the ribbon.
Note that the integral equation in Eq.~\eqref{6:SREs} is dimensionless; lengths have been scaled   by $\ell$, velocities by a typical ribbon velocity $U$, forces by $\mu \ell U$ and torques by $\mu \ell^{2} U$.   Furthermore, in order to obtain Eq.~\eqref{6:SREs} one assumes that   $\rho(s_{1})$ is locally ellipsoidal near the ends of the ribbon. 
 
 The total force and torque on the fluid from the ribbon are then given by
\begin{eqnarray}
\mathbf{F}_{h} &=& \int_{-1}^{1} \,ds_{1} \int_{-1}^{1} \,ds_{2} \mbox{ } \mathbf{f}(s_{1},s_{2}), \\
\mathbf{L}_{h} &=& \int_{-1}^{1} \,ds_{1} \int_{-1}^{1} \,ds_{2} \mbox{ } \mathbf{Y}(s_{1},s_{2})\times \mathbf{f}(s_{1},s_{2}),
\end{eqnarray}
where $\mathbf{Y}(s_{1},s_{2}) = \mathbf{r}(s_{1}) + b_{\ell} s_{2} \rho(s_{1}) \mathbf{\hat{T}}(s_{1})$ is the scaled ribbon plane. These equations have been shown to accurately capture both known theoretical results and experimental measurements  \cite{Koens2016}.

 \subsection{Looped ribbons}
 The above equations characterise the hydrodynamics of a finite ribbon of total length $2\ell$. The extension  to  looped ribbons is found through a similar derivation to that of Ref.~\cite{Koens2016}, but with $s'_{1}$ replaced by a $s_{1}+q$ where $q$ is now the integration variable (see details in Appendix~\ref{sec:a2}). This substitution describes a looped system as the integration becomes independent of the choice of origin ($s_{1}=0$). As shown in Appendix~\ref{sec:a2}, the SRT equations for looped ribbons are equivalent to substituting
\begin{eqnarray}
s'_{1} &\rightarrow & s_{1} +q, \\
\int_{-1}^{1} \,ds'_{1} &\rightarrow & \int_{-1}^{1} \,dq, \\
\ln\left(\frac{4(1-s_{1}^{2})}{b_{\ell}^{2} \rho^{2}(s_{1}) (s_{2}-s'_{2})^{2}}\right)  &\rightarrow & \ln\left(\frac{4}{b_{\ell}^{2} \rho^{2}(s_{1}) q^{2}}\right),
\end{eqnarray}
into Eq.~\eqref{6:SREs}, with the understanding that $\rho(s_{1})$ remains non-zero anywhere along the centreline.

\section{Analytical solutions}

\subsection{The potential difficulty} \label{diff}

Due to the first and second integrals on the right hand side of Eq.~\eqref{6:SREs}, it is unclear if the SRT integral equation  has any rigid-body analytical solutions. The first integral, which we term   the outer integral, closely resembles  the outer integral in slender-body theory (integral in Eq.~\ref{SBT}). In the case of slender bodies, this integral can be simplified for simple shapes such as rods \cite{Johnson1979,Tornberg2006} and tori \cite{Johnson1979a}, and an analogous simplification is probably doable for ribbons as well. 

The second integral on the right-hand side of Eq.~\eqref{6:SREs}, which we call the logarithm integral, is however new to the SRT equations.  This integral is done over the width of the ribbon, only depends on the scaled-ribbon plane locally (i.e.~it is independent of $s'_{1}$) \cite{Koens2016}, and is the only term involving $s_{2}$ on the right hand side of Eq.~\eqref{6:SREs}. As a consequence, any velocity of the surface of the ribbon with non-zero $s_{2}$ dependence is generated from this integral.  The requirement to generate the $s_{2}$ motion therefore determines  the force distribution across the ribbons width (the $s'_{2}$ dependence) for all ribbons. Splitting  the logarithm integral as
\begin{eqnarray}
 &&\int_{-1}^{1} \,d s'_{2} \ln\left(\frac{4(1-s_{1}^{2})}{b_{\ell}^{2} \rho^{2} (s_{2}-s'_{2})^{2}}\right)  \left(\mathbf{I} +\mathbf{\hat{t}} \mathbf{\hat{t}} \right) \cdot  \mathbf{f}(s_{1},s'_{2}) \notag \\ 
 &&=  \ln\left(\frac{4(1-s_{1}^{2})}{b_{\ell}^{2} \rho^{2}}\right)  \left(\mathbf{I} +\mathbf{\hat{t}} \mathbf{\hat{t}} \right) \cdot   \left\langle \mathbf{f} \right\rangle (s_{1})
+ \left(\mathbf{I} +\mathbf{\hat{t}} \mathbf{\hat{t}} \right) \cdot  \int_{-1}^{1} \,d s'_{2} \ln\left(\frac{1}{(s_{2}-s'_{2})^{2}}\right)    \mathbf{f}(s_{1},s'_{2}),
\end{eqnarray}
explicitly separates the behaviour which depends on $s_{2}$ (second term) to that without (first term). Since this second integral, in combination with the $s_{2}$ dependence of the velocity, defines the force distribution in $s_{2}'$ to within an arbitrary proportionality constant we can focus on the integral
\begin{equation} \label{6integral}
I (s_1,s_2)= \int_{-1}^{1} \,d s'_{2} \ln\left(\frac{1}{(s_{2}-s'_{2})^{2}}\right) \mathbf{f}(s_{1},s'_{2}),
\end{equation}
instead of the full logarithm integral without any loss of generality. This above integral has no dependence on the scaled-ribbon plane, indicating that the force distribution in $s'_2$ is independent of the ribbon's shape. Hence using $I(s_1,s_2)$ the force distributions in $s_{2}'$ can be determined generally and then inserted into Eq.~\eqref{6:SREs} to simplify the general slender-ribbon equations. 

\subsection{The force distribution along the width (in $s_{2}$)} \label{fdis}

Analytical solutions to SRT require knowledge of the  dependence of the force density across the width of the ribbon, i.e.~along the $s_2$ direction.  As discussed above, this distribution is independent of ribbon's shape and when inserted into Eq.~\eqref{6integral} it produces the  $s_{2}$ dependence of ribbon's velocity (i.e.~the left-hand side of Eq.~\ref{6:SREs}).  It is therefore important to determine how the velocity of the ribbon depends on $s_{2}$ for an arbitrary motion and deformation.

One of the important underlying assumptions of SRT is that the surface of the slender ribbon moves rigidly with the scaled-ribbon plane. Since a scaled-ribbon plane undergoing an arbitrary deformation is described by
\begin{equation}
\mathbf{Y}(s_{1},s_{2},t) = \mathbf{r}(s_{1},t) + b_{\ell} s_{2} \rho(s_{1}) \mathbf{\hat{T}}(s_{1},t),
\end{equation}
the surface velocity of a slender ribbon undergoing rigid-body translation at speed $\mathbf{U}_{r}$ and angular rotation at speed $\boldsymbol{\Omega}_{r}$ becomes
\begin{eqnarray} \label{surfvel}
\mathbf{U}(s_{1},s_{2},t) &=& \mathbf{U}_{r} + \boldsymbol{\Omega}_{r}\times \mathbf{Y}(s_{1},s_{2},t) + \partial_{t} \mathbf{Y}(s_{1},s_{2},t)\notag  \\
&=& \mathbf{U}_{r} + \boldsymbol{\Omega}_{r}\times\mathbf{r}(s_{1},t) + \partial_{t} \mathbf{r}(s_{1},t) + b_{\ell} s_{2} \rho(s_{1}) \left[\boldsymbol{\Omega}_{r}\times \mathbf{\hat{T}}(s_{1},t) + \partial_{t} \mathbf{\hat{T}}(s_{1},t)  \right], 
\end{eqnarray}
where $t$ denotes time. 
 The above equation shows that, for an arbitrary deformation and rigid-body motion, the ribbon's velocity is at most linear in $s_{2}$.
 With this in mind we may redefine the force as
\begin{equation}
\mathbf{f} = \mathbf{f}_{1} (s_{1}) g_{1}(s_{2}) +\mathbf{f}_{2} (s_{1}) g_{2}(s_{2}), \label{force}
\end{equation}
where the functions $g_{1}(s_{2})$ and $g_{2}(s_{2})$ satisfy
\begin{eqnarray}
\int_{-1}^{1} \ln\left(\frac{1}{(s'_{2}-s_{2})^{2}} \right) g_{1}(s'_{2}) \,ds'_{2} &=& 2 \pi \ln(2), \\
\int_{-1}^{1} \ln\left(\frac{1}{(s'_{2}-s_{2})^{2}} \right) g_{2}(s'_{2}) \,ds'_{2} &=& 2 \pi s_{2}.
\end{eqnarray}
 In the above $\mathbf{f}_{1} (s_{1})$ represents the force distribution along the centreline generated from motions that do not involve the width arclength $s_{2}$, $\mathbf{f}_{2} (s_{1})$ represents the force distribution along the centreline from motions that involve $s_{2}$ linearly, and the proportionality constants of the integral equations where chosen for future simplicity.
 
 The above integral equations are special cases of Carleman's equation \cite{Carleman1922,Polianin2008}. Carleman showed that for integral equations of the form
 \begin{equation}
\int_{-1}^{1} \ln\left|s'_{2}-s_{2} \right| g(s'_{2}) \,ds'_{2} = f(s_{2}),
 \end{equation}
 $g(s'_{2})$ can be written in terms of integrals of $f(s_{2})$, where $g(s'_{2})$ is the unknown function and $f(s_{2})$ is the arbitrary forcing \cite{Carleman1922}. These general integrals are listed Ref.~\cite{Polianin2008}, Eqs.~3.4.2-4,  and can easily evaluated in the case of linear or constant $f(s_{2})$. Hence, using these results, the $g_{1}$ and $g_{2}$ distributions are
\begin{eqnarray}
g_{1}(s_{2}) &=& \frac{1}{\sqrt{1-s_{2}^{2}}}, \label{g1}\\
g_{2}(s_{2}) &=& \frac{s_{2}}{\sqrt{1-s_{2}^{2}}}\cdot \label{g2}
\end{eqnarray} 
The above $1/\sqrt{1-s^{2}}$ force dependence is  likely a result of taking the asymptotically thin limit of the ribbons surface, $\ell \gg b \gg a$. Though this dependence gives an infinite force density at the edges, this force distribution lies on an imaginary plane within the ribbon and therefore no actual point over the  surface of the ribbon experiences this force. Furthermore, the total force across the ribbons width, $s_{2}$, is finite and so measurable values of the  force and moments are finite. A similar divergence is seen for  an infinitely thin flat plate in potential flow where the velocity profile  has a $s/\sqrt{1-s^{2}}$ velocity distribution along the plates surface \cite{White2003}. 

\section{The reduced slender-ribbon theory  equations} \label{rSRT}

In the previous section, we determined that the force distribution across a ribbon's width can be written as Eqs.~\eqref{g1} and \eqref{g2}. These functions are integrable and simplify the logarithmic integral, Eq.~\eqref{6integral}, to generate the relevant ribbon motion. The $g_{i}(s_{2})$ functions can therefore be used to considerably simplify the SRT equations for an arbitrary isolated ribbon.

Inserting Eqs.~\eqref{force}, \eqref{g1} and \eqref{g2}  into Eq.~\eqref{6:SREs} the equations for an arbitrary isolated slender-ribbon reduce to
\begin{eqnarray}\label{eq:newSRT}
\notag 8 \pi \mathbf{U}(s_1,s_2) &=& \pi \int_{-1}^{1} \,d s'_{1}  \left[\frac{ (\mathbf{I} +\mathbf{\hat{R}}_{0} \mathbf{\hat{R}}_{0} ) \cdot \mathbf{f}_{1}(s'_{1})}{|\mathbf{R}_{0}|}-\frac{ (\mathbf{I} +\mathbf{\hat{t}} \mathbf{\hat{t}} ) \cdot \mathbf{f}_{1}(s_{1})}{|s'_{1}-s_{1}|}\right] \\
\notag &&+  \pi \left[ L_{SRT} (\mathbf{I} +\mathbf{\hat{t}} \mathbf{\hat{t}}) 
-2 \mathbf{\hat{t}} \mathbf{\hat{t}} + 2 \mathbf{\hat{T}} \mathbf{\hat{T}}  \right] \cdot \mathbf{f}_{1}(s_{1}) \\&&
+  2 \pi (\mathbf{I} +\mathbf{\hat{t}} \mathbf{\hat{t}}) \cdot \left( \ln(2)  \mathbf{f}_{1} (s_{1}) +  s_{2} \mathbf{f}_{2} (s_{1})\right), \label{SRTred}
\end{eqnarray}
for finite bodies with $\rho=\sqrt{1-s_{1}^{2}}$ or 
\begin{eqnarray} 
\notag 8 \pi \mathbf{U}(s_1,s_2) &=& \pi \int_{-1}^{1} \,d q \left[\frac{ (\mathbf{I} +\mathbf{\hat{R}}_{0} \mathbf{\hat{R}}_{0} ) \cdot \mathbf{f}_{1}(s_{1}+q)}{|\mathbf{R}_{0}|}-\frac{ (\mathbf{I} +\mathbf{\hat{t}} \mathbf{\hat{t}} ) \cdot \mathbf{f}_{1}(s_{1})}{|q|}\right] \\
\notag &&+  \pi \left[ L_{SRT} (\mathbf{I} +\mathbf{\hat{t}} \mathbf{\hat{t}}) 
-2 \mathbf{\hat{t}} \mathbf{\hat{t}} + 2 \mathbf{\hat{T}} \mathbf{\hat{T}}  \right] \cdot \mathbf{f}_{1}(s_{1}) \\&&
+ 2 \pi (\mathbf{I} +\mathbf{\hat{t}} \mathbf{\hat{t}}) \cdot \left( \ln(2)  \mathbf{f}_{1} (s_{1}) + s_{2} \mathbf{f}_{2} (s_{1})\right), \label{SRTring}
\end{eqnarray}
for looped bodies. In these equations we have denoted $L_{SRT} = \ln(4/b_{\ell}^{2})$. 

Comparing Eq.~\eqref{6:SREs} with the SRT equations above, Eq.~\eqref{eq:newSRT}, we see that it has now been reduced from a series of line and surface integrals into a single line integral with additional  constant terms.  Not only is this structure  very similar to the slender-body theory equations, the line integral is identical to the line integral in slender-body theory, Eq.~\eqref{SBT}, with an added pre-factor of $\pi$. This correspondence allows any centreline, $\mathbf{r}(s_{1})$, previously calculated using slender-body theory to be easily adapted to the case of slender ribbons. Furthermore, since  slender-body theory is known to possess analytical solutions in the case of rigid motions, we expect analytical slender-ribbon analogues to also exist.

\section{Rigid-body analytical solutions to slender-ribbon theory} \label{analytic}

There exists two classic analytical solutions to slender-body theory: the thin prolate ellipsoid \cite{Johnson1979,Tornberg2006} and the cylindrical torus \cite{Johnson1979a}. 
 In this section we take advantage of the correspondence between the SBT and SRT equations to characterise theoretically the rigid-body motion of  long flats ellipsoids and of a ribbon torus.

\subsection{Rigid-body motion of a long flat ellipsoid} \label{arod}

\begin{figure}[t]
\centering
\includegraphics[width=0.8\textwidth]{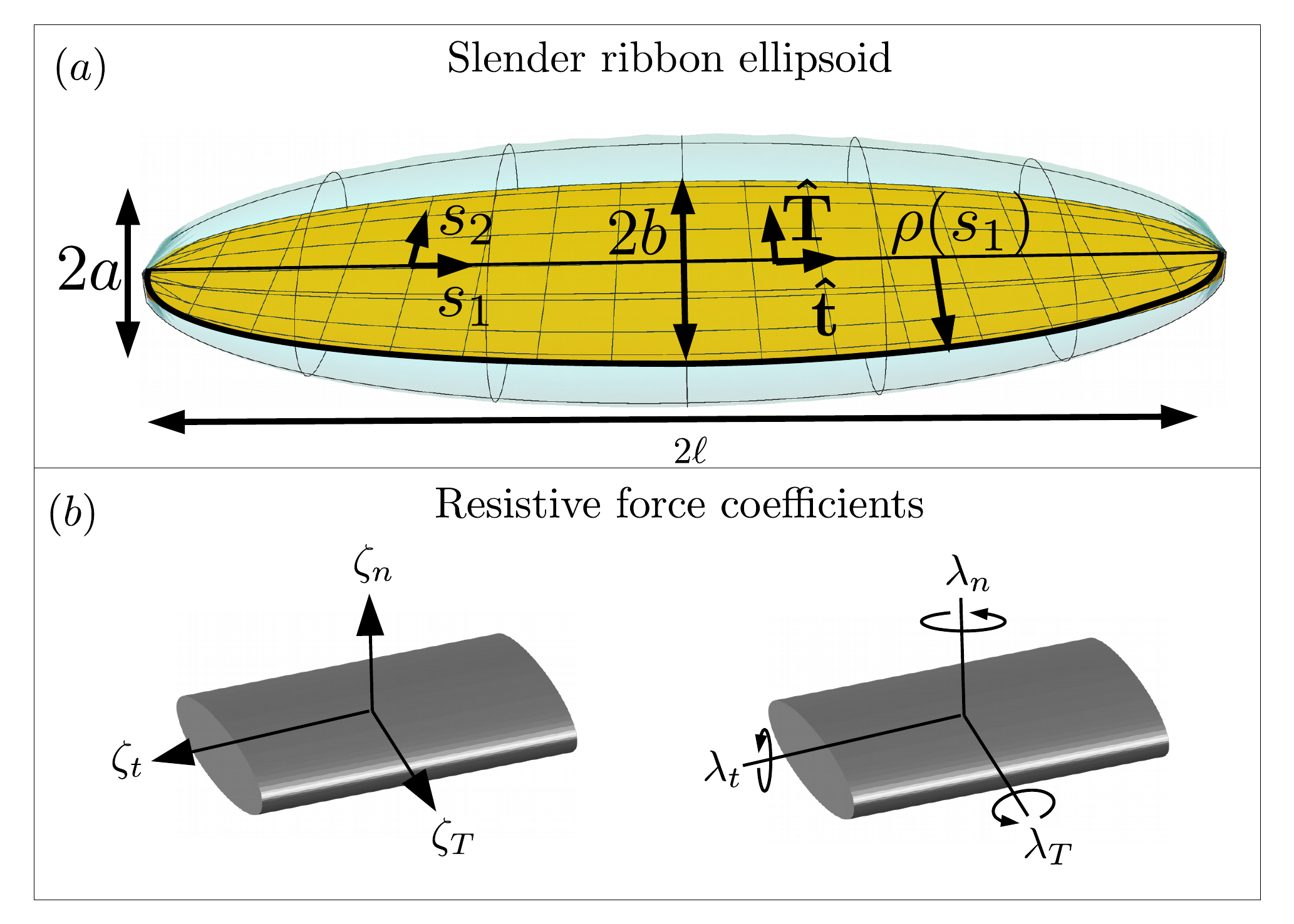}
\caption{(a) A long flat ellipsoid with all the parameters describing the ribbon illustrated; (b) Image depicting the ribbon resistance coefficients: translation (left) and rotation (right).}
\label{fig:rodconfig}
\end{figure}

The long flat ellipsoid is the simplest shape that slender-ribbon theory can describe. In this case, the  centreline is  straight and the vector $\mathbf{\hat{T}}$ is constant. Formally this ellipsoidal structure is given by
\begin{eqnarray}
\mathbf{r}(s_{1}) &=& s_{1} \mathbf{\hat{x}}, \\
\mathbf{\hat{T}} &=& \mathbf{\hat{y}}.
\end{eqnarray}
This shape illustrated in Fig.~\ref{fig:rodconfig}a. 
With this parametrisation the remaining integral in Eq.~\eqref{SRTred} has eigenfunctions of Legendre polynomials with known eigenvectors \cite{Gotz2000, Koens2016, Koens2014}. In addition, only the zeroth ($P_{0}(s_{1})=1$) and first ($P_{1}(s_{1})=s_{1}$) Legendre polynomial will be needed to solve these equations for rigid motion.

\subsubsection{Translation}

We first consider the rigid translation of the long flat ellipsoid. For all rigid translations the velocity is constant across the sheet. Therefore  $\mathbf{f}_{1}$ should be constant and $\mathbf{f}_{2}(s_{1}) = {\bf 0}$. The equations to solve then are
\begin{equation}
 8 \pi \mathbf{U} =  \pi \left[ (L_{SRT}+\ln(4)) (\mathbf{I} +\mathbf{\hat{x}} \mathbf{\hat{x}}) 
-2 \mathbf{\hat{x}} \mathbf{\hat{x}} + 2 \mathbf{\hat{y}} \mathbf{\hat{y}}  \right] \cdot \mathbf{f}_{1}.
\end{equation}
The total force, $\mathbf{F}_{h}$,  and torque, $\mathbf{L}_{h}$,  acting on the fluid (i.e.~opposite to the drag) as a result of the rigid-body translation of a long flat ellipsoid are thus
\begin{eqnarray}
\mathbf{F}_{h} &=& 16 \pi \left[ (L_{SRT}+\ln(4)) (\mathbf{I} +\mathbf{\hat{x}} \mathbf{\hat{x}}) 
-2 \mathbf{\hat{x}} \mathbf{\hat{x}} + 2 \mathbf{\hat{y}} \mathbf{\hat{y}}  \right]^{-1} \mathbf{U} , \\
\mathbf{L}_{h} &=& \mathbf{0}.
\end{eqnarray}
The above force exhibits a structure $\propto 1/(\ln(\ell/b)+\xi)$, where $\xi$ is some constant, very similar to  the forces exerted by  a thin prolate ellipsoid, $\propto 1/(\ln(2\ell/b) \pm 1/2)$ \cite{Chwang2006}. These coefficients are identical to previously calculated drag coefficients for an ellipsoid using SRT \cite{Koens2016}(not shown).

\subsubsection{Rotation}

The hydrodynamics of rigid rotation is now considered for plate ellipsoids. In the case of rigid-body rotation the velocity is proportional to $s_{1}$ and $s_{2}$. Hence $\mathbf{f}_{1}(s_{1}) = s_{1}\mathbf{c}_{1}$  and $\mathbf{f}_{2}(s_{1}) \neq 0$, where $\mathbf{c}_{1}$ is an unknown constant vector. Separating the constant terms from those  proportional  $s_{2}$, the equations become
\begin{eqnarray}
 8 \pi s_{1} \boldsymbol{\Omega}\times\mathbf{\hat{x}} &=& \pi s_{1} \left[ (L_{SRT} +\ln(4) -2)(\mathbf{I} +\mathbf{\hat{x}} \mathbf{\hat{x}}) 
-2 \mathbf{\hat{x}} \mathbf{\hat{x}} + 2 \mathbf{\hat{y}} \mathbf{\hat{y}}  \right] \cdot \mathbf{c}_{1},\\
8 \pi b_{l} \rho(s_{1}) s_{2} \boldsymbol{\Omega}\times\mathbf{\hat{y}} &=&  2 \pi s_{2} (\mathbf{I} +\mathbf{\hat{x}} \mathbf{\hat{x}}) \cdot \mathbf{f}_{2} (s_{1}).
\end{eqnarray}
thereby providing the body with a net force and torque of
\begin{eqnarray}
\mathbf{F}_{h} &=& \mathbf{0} , \\
\notag\mathbf{L}_{h} &=& \frac{16 \pi}{3} \mathbf{\hat{x}}\times \left\{\left[ (L_{SRT} +\ln(4) -2)(\mathbf{I} +\mathbf{\hat{x}} \mathbf{\hat{x}}) 
-2 \mathbf{\hat{x}} \mathbf{\hat{x}} + 2 \mathbf{\hat{y}} \mathbf{\hat{y}}  \right]^{-1} \boldsymbol{\Omega}\times\mathbf{\hat{x}}\right\} \\
&& +\frac{8 \pi b_{l}^{2} }{3} \mathbf{\hat{y}}\times\left[ (\mathbf{I} +\mathbf{\hat{x}} \mathbf{\hat{x}})^{-1}\boldsymbol{\Omega}\times\mathbf{\hat{y}}\right].
\end{eqnarray}
 Again these coefficients agree with the previously numerically calculated resistance coefficients \cite{Koens2016}.
Also we note that the final term of the torque is very small; however this term  was shown to be accurate numerically in Ref.~\cite{Koens2016}.

\subsubsection{The resistance matrix and a ribbon resistive-force theory}

The resistance coefficients of  slender-bodies with ellipsoidal cross sections has been investigated previously by Batchelor \cite{Batchelor2006}. This was done by using stokeslets to derive an integral equation for the force, accurate to order $b_{\ell} \ln(b_{\ell})$, and then solving this equation iteratively in powers $1/\ln(b_{\ell})$. In the ellipsoidal limit this expansion could then be solved to order $b_{\ell} \ln(b_{\ell})$. This gave an ellipsoid  with semi-axes lengths 1, $b_{\ell}$ and $a_{\ell}$ a resistance matrix, $\mathbf{R}_{B}$, of
\begin{eqnarray}\label{eq:Batchelor}
\mathbf{R}_{B} =  \left(\begin{array} {c c c c c c}
\frac{8 \pi }{2 \ln\left[ \frac{4}{b_{\ell}+a_{\ell}}\right] -1} &0 &0 & 0 &0 &0\\
0& \frac{8 \pi }{ \ln\left[ \frac{4}{b_{\ell}+a_{\ell}}\right] +\frac{b_{\ell}}{b_{\ell}+a_{\ell}}}& 0 & 0 &0 &0 \\
0& 0&  \frac{8 \pi }{ \ln\left[ \frac{4}{b_{\ell}+a_{\ell}}\right]+\frac{b_{\ell}}{b_{\ell}+a_{\ell}}} & 0 &0 &0\\
0&0&0& 0&0 &0 \\
0&0&0&0& \frac{8\pi }{3\left( \ln\left[ \frac{4}{b_{\ell}+a_{\ell}}\right] -\frac{b_{\ell}}{b_{\ell}+a_{\ell}}\right)}& 0 \\
0&0&0&0& 0&  \frac{8\pi }{3\left( \ln\left[ \frac{4}{b_{\ell}+a_{\ell}}\right] -\frac{b_{\ell}}{b_{\ell}+a_{\ell}}\right)}
\end{array}  \right). \notag \\
\end{eqnarray}
The results in Eq.~\eqref{eq:Batchelor} are identical to our SRT analytical solutions in the limit $a_{\ell} \rightarrow 0$, to $O(b_{\ell})$, thereby confirming our results.

With these results, a resistive-force theory for ribbons (RRFT) can be proposed. These theories are practical, as they can provide physical insight and analytical approximations of the drag and dynamics of a system \cite{GRAY1955,Lauga2009}. Fundamentally resistive-force theories assume that the local hydrodynamics of any point along a slender body is similar to a the dynamics of a straight body with the same cross section. As a result the force and torque per unit length, at given point on the body, is approximately equal to the force and torque per unit length experienced by a straight body for the same motion. This produces a linear relationship between the local force and motion of the body. In this classic derivation for slender cylindrical filaments (RFT) \cite{GRAY1955,Lauga2009}, the asymmetric cross section creates two proportionality coefficients, however for ribbons three coefficients are needed to capture the three dimensional cross sectional shape. Therefore the local force and torque can be written as
\begin{eqnarray}
\mathbf{f}_{RRFT}(s_{1}) &=& \left[ \zeta_{t}^{RFT}\mathbf{\hat{t}} \mathbf{\hat{t}} +  \zeta_{T}^{RFT} \mathbf{\hat{T}} \mathbf{\hat{T}} +  \zeta_{n}^{RFT} (\mathbf{\hat{t}}\times\mathbf{\hat{T}}) (\mathbf{\hat{t}}\times \mathbf{\hat{T}})\right] \cdot \mathbf{U} , \label{frrft} \\
\boldsymbol{\ell}_{RRFT}(s_{1}) &=& \left[\lambda_{t}^{RFT} \mathbf{\hat{t}} \mathbf{\hat{t}} +  \lambda_{T}^{RFT} \mathbf{\hat{T}} \mathbf{\hat{T}} +  \lambda_{n}^{RFT} (\mathbf{\hat{t}}\times\mathbf{\hat{T}}) (\mathbf{\hat{t}}\times \mathbf{\hat{T}})\right] \cdot \boldsymbol{\Omega}, \label{lrrft}
\end{eqnarray}
where $\mathbf{f}_{RRFT}$ is the approximate force per unit length at $s_{1}$, $ \boldsymbol{\ell}_{RRFT}$ is the approximate torque per unit length at $s_{1}$, $\zeta_{t}^{RFT}$, $ \zeta_{T}^{RFT}$, and $\zeta_{n}^{RFT}$ are the local resistance coefficients relating the force to the linear velocity at $s_{1}$, and $\lambda_{t}^{RFT}$, $ \lambda_{T}^{RFT}$, and $\lambda_{n}^{RFT}$ are the local resistance coefficients relating the torque and angular velocity at $s_{1}$. The subscripts on the resistance coefficients denote their directionality, subscript $t$ relating to the tangent direction, subscript $T$ relating to the $\mathbf{\hat{T}}$ direction and subscript $n$ relating to the normal direction ($\mathbf{\hat{t}}\times\mathbf{\hat{T}}$). The total force and torque on a body, in RRFT, is therefore given by
\begin{eqnarray}
\mathbf{F}_{RRFT} &=& \int_{-1}^{1} \,ds_{1} \mathbf{f}_{RRFT}(s_{1}), \\
\mathbf{L}_{RRFT} &=& \int_{-1}^{1} \,ds_{1} \left[ \mathbf{r}(s_{1}) \times\mathbf{f}_{RRFT}(s_{1}) + \boldsymbol{\ell}_{RRFT}(s_{1}) \right],
\end{eqnarray}
where $\mathbf{F}_{RRFT} $ and $\mathbf{L}_{RRFT}$ are the total force and torque, respectively. These resistive-force theories require the body to be exponentially thin. This is because the resistance from the long range hydrodynamics interactions, present in the outer expansion region, is $\ln(b_{\ell})$ smaller than the terms found in the inner region. The logarithms in Eqs.~\eqref{SBT}, \eqref{SRTred}, \eqref{SRTring} are a manifestation of this. As a result it is typically used to only capture the governing physics qualitatively.

To determine the local resistance coefficients for RRFT we refer to the force and torque distributions found for a long flat ellipsoid. Since RRFT assumes that any point is a locally straight ribbon, the force an torque per unit length experienced by a point is therefore equivalent to the force and torque per unit length of the long flat ellipsoid at its center, $s_{1}=0$. Therefore by integrating the force and torque distributions at $s_{1}=0$ over $s_{2}$, and comparing the resultant drag with the structure of Eqs.~\eqref{frrft} and \eqref{lrrft}, we find
\begin{eqnarray}
\zeta_{t}^{RFT} &=& \frac{4 \pi }{2 \ln( 4/b_{\ell}) -1}, \\
\zeta_{T}^{RFT} &=&  \frac{4 \pi }{ \ln( 4/b_{\ell}) +1},\\
\zeta_{n}^{RFT} &=& \frac{4 \pi }{ \ln( 4/b_{\ell})}, \\
\lambda_{t}^{RFT} &=& 2 \pi b_{\ell}^{2}, \\
\lambda_{T}^{RFT} &=& 0,\\
\lambda_{n}^{RFT} &=&  \pi b_{\ell}^{2}.
\end{eqnarray}
Note that the classic resistive-force theory for slender filaments does not include a torque relation equivalent to Eq.~\eqref{lrrft}. This is due to these coefficients typically being negligible. We however have included it here for the sake of completeness. Furthermore, it is possible to modify these coefficients to handle ellipsoidal  cross sections using the results of Batchelor \cite{Batchelor2006}, without any loss of generality.

\subsubsection{The sedimentation of a long flat ellipsoid}

The resistance coefficients for the long flat ellipsoid allow us to consider how these shapes sediment under the action of gravity. It is well known that anisotropic bodies, such as rods, settle in general at an angle to the applied gravitational force, which we term deflection angle. In his famous ``Low-Reynolds-number flows'' movie, G.~I.~Taylor showed that the maximum deflection angle for rods, with drag coefficient perpendicular to the road  twice the parallel drag coefficient, was approximately $19^{\circ}$ \cite{Taylor1967}.

\begin{figure}[t]
\centering
\includegraphics[width=.45\textwidth]{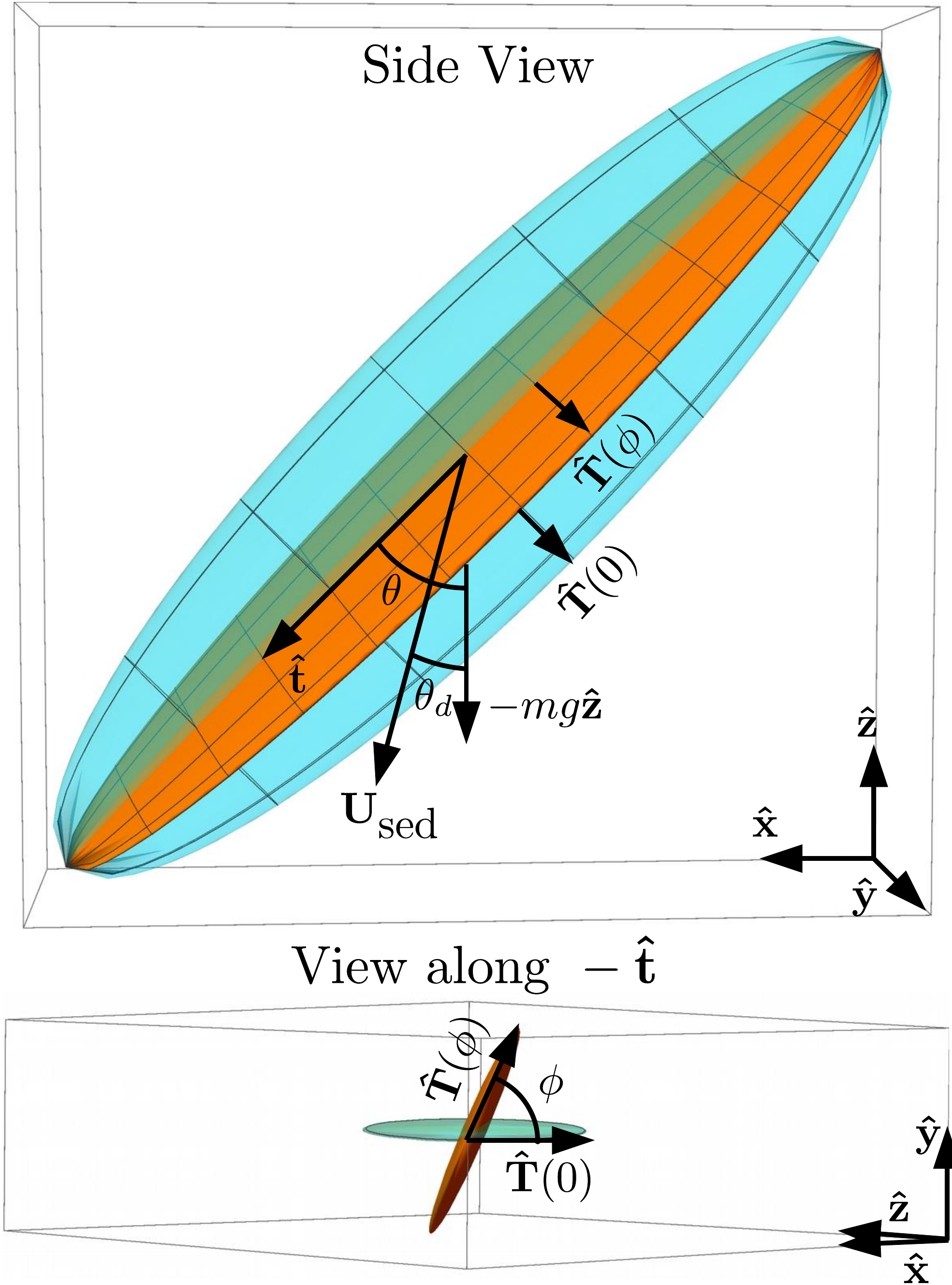}
\caption{A settling long flat ellipsoid. Top: Side view;  Bottom: Ellipsoid viewed down $-\mathbf{\hat{t}}$; $\theta$ is the angle between the centreline of the ellipsoid and the gravitational force, $\phi$ is the angle between $\mathbf{\hat{T}}$ and the plane of the force and centerline and $\theta_{d}$ is the angle between the sedimentation velocity and the gravitation force. The blue translucent image shows the ellipsoid with $\phi=0$, while the orange solid image shows the ellipsoid when $\phi \neq 0$.}
\label{fig:sedimentrod}
\end{figure}

The sedimentation velocity of a body  of mass $m$ can be found by balancing the gravitational force, $-mg\mathbf{\hat{z}}$, with the hydrodynamic forces. The flat ellipsoid under gravity is then described by two angles: $\theta$ which measures the angle between the centreline of the ellipsoid and the gravitational force and $\phi$ which gives the angle of the ribbons width, the vector $\mathbf{\hat{T}}$, to the plane of the force and centreline (see sketch in Fig.~\ref{fig:sedimentrod}). For convenience we define the plane with the force and centreline to be in the $\mathbf{\hat{z}}$-$\mathbf{\hat{x}}$ plane. The sedimentation velocity, $\mathbf{U}_{sed}$, of a long flat ellipsoid with length $ 2\ell$ then is
\begin{equation}
\mathbf{U}_{sed} = -\frac{mg}{2  \mu \zeta_{t} \zeta_{T} \zeta_{n} \ell} \left(\begin{array}{c}
-\frac{1}{4} \sin(2 \theta) \left[\zeta_{t} (\zeta_{T} +\zeta_{n})-2 \zeta_{T} \zeta_{n}+\zeta_{t}(\zeta_{n}-\zeta_{T}) \cos(2 \phi) \right] \\
- \zeta_{t} \left[ (\zeta_{n}-\zeta_{T}) \sin\theta \cos(\phi) \sin(\phi) \right] \\
\zeta_{T} \zeta_{n} \cos^{2}(\theta) + \zeta_{t} \sin^{2}(\theta)\left[\zeta_{n} \cos^{2}(\phi) +\zeta_{T} \sin^{2}(\phi) \right]
\end{array} \right),
\end{equation}
in the $(x,y,z)$ frame.

The deflection angle, $\theta_{d}$, is defined as the angle between $\mathbf{U}_{sed}$ and $-mg\mathbf{\hat{z}}$ and is solution to
\begin{equation}
\tan^{2}(\theta_{d}) = \frac{U_{x}^{2} + U_{y}^{2}}{U_{z}^{2}},
\end{equation}
where $U_{i}$ is the velocity component in direction $i$. From this equation the maximum value of  $\theta_{d}$ can be found by maximising the right hand side with respect to $\theta$ and $\phi$. Since $\zeta_{n}>\zeta_{T}>\zeta_{t}$, $U_{x}^{2} + U_{y}^{2}$ is maximised, and $U_{z}^{2}$ is minimised, for $\phi=\pi/2$, irrespective of $\theta$. The maximum deflection occurs therefore  when the motion is two dimensional and depends only on $\zeta_{t}$ and $\zeta_{n}$. This motion is identical to a settling rod and so the maximum deflection angle, between the settling velocity and gravity, is given by
\begin{equation}
\tan(\theta_{d}^{\mbox{max}}) = \frac{\zeta_{t} - \zeta_{n}}{2 \sqrt{\zeta_{n} \zeta_{t}}},
\end{equation}
when the angle between the ellipsoid's centerline and gravity, $\theta$, satisfies $\cos(2\theta)=(\zeta_{t}-\zeta_{n})/(\zeta_{t}+\zeta_{n})$ \cite{Guyon2001}.  This orientation maximises the deflection as it creates the largest difference between the drag parallel and perpendicular to gravity with the three resistance coefficients available.

\subsection{Rigid-body motion of a ribbon torus} \label{atorus}

After having addressed rigid ellipsoids, we now consider  another shape for which slender-ribbon theory
 can provide analytical solutions, namely the ribbon torus. Indeed, such a shape has circular symmetries which can be exploited to simplify the outer integral. 
 \begin{figure}
\centering
\includegraphics[width=0.6\textwidth]{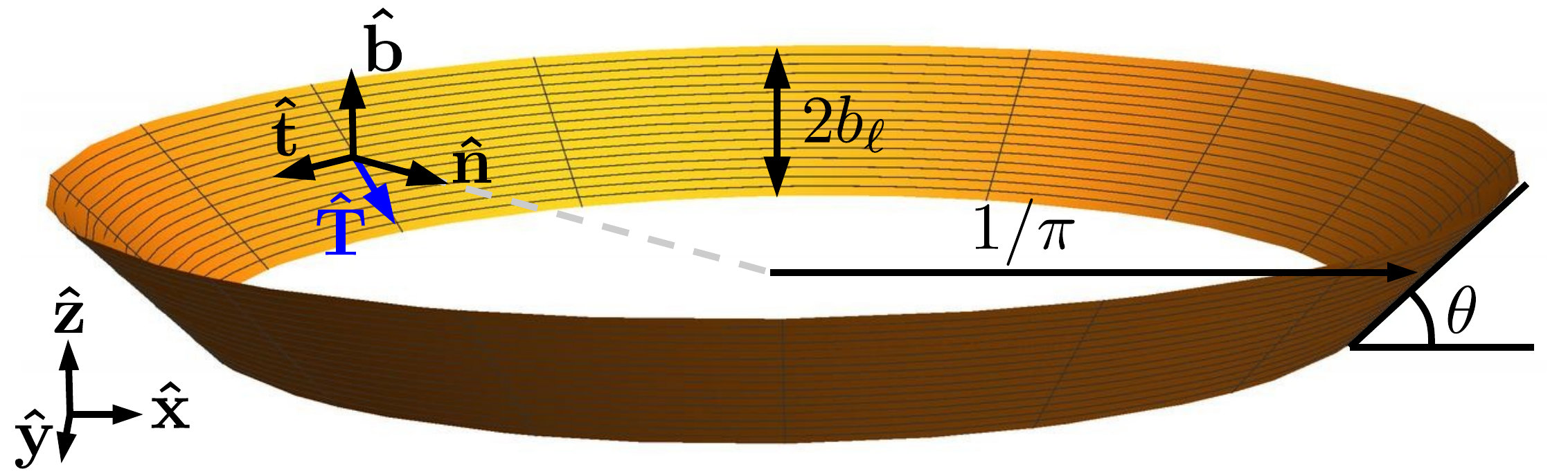}
\caption{The scaled-ribbon plane of a ribbon torus: $\mathbf{\hat{t}}$ is the tangent vector to the centreline, $\mathbf{\hat{n}}$ is the normal vector to the centreline, $\mathbf{\hat{b}}$ is the bi-normal vector, and $\mathbf{\hat{T}}$ is a vector pointing towards the major axis of the ribbon's cross section. This torus has a width of $2 b_{\ell}$, a radius of $1/\pi$ and 
the angle between the ribbons plane and the centerline's normal vector is denoted $\theta$.}
\label{fig:torusconfig}
\end{figure}

  The shape of a ribbon torus, illustrated in Fig.~\ref{fig:torusconfig},  is mathematically described by
\begin{eqnarray}
\mathbf{r} (s_{1})&=& \frac{1}{\pi}\left\{\cos(\pi s_{1}),\sin(\pi s_{1}),0  \right\}, \\
\mathbf{\hat{t}}(s_{1}) &=& \left\{-\sin(\pi s_{1}),\cos(\pi s_{1}),0  \right\}, \\
\mathbf{\hat{T}}(s_{1}) &=& \cos\left(\theta\right)\mathbf{\hat{n}}(s_{1}) +\sin\left(\theta\right)\mathbf{\hat{b}},
\end{eqnarray}
where $\mathbf{\hat{n}}(s_{1})=\{-\cos(\pi s_{1}),-\sin(\pi s_{1}),0\}$ is the normal vector to the centreline and $\mathbf{\hat{b}}= \{0,0,1\}$ is the bi-normal vector to the centreline. In this parametrisation $\theta$ determines how the ribbon sits relative to the $\mathbf{r}$ plane: When $\theta=0$ the ribbon lies completely in the $\mathbf{r}$ plane, while when $\theta=\pi/2$ the ribbon sits perpendicular to it. The circular symmetries of this shape prompts us to divide the force into components along $\mathbf{\hat{t}}$, $\mathbf{\hat{n}}$ and $\mathbf{\hat{b}}$, and to write
\begin{equation}
\mathbf{f}_{i} = f_{i,t}(s_{1}) \mathbf{\hat{t}} +f_{i,n}(s_{1}) \mathbf{\hat{n}} +f_{i,b}(s_{1}) \mathbf{\hat{b}}.
\end{equation}
This parametrisation allows the rigid-body hydrodynamics of a ribbon torus to be found from Eq.~\eqref{SRTring}. Furthermore $\mathbf{f}_{2}$ is neglected for all these calculations as it always of order $b_{\ell}$ or higher. This due to all the terms proportional to $s_{2}$ in the surface velocity, Eq.~\eqref{surfvel}, also being proportional to $b_{\ell}$. Since the system is linear and $\mathbf{f}_{2}$ must account for the terms proportional to $s_{2}$, $\mathbf{f}_{2}$ must also be proportional to $b_{\ell}$, thereby making it negligible. 

\subsubsection{Translation along torus axis ($\mathbf{\hat{z}}$)}

\begin{figure}[t]
\centering
\includegraphics[width=0.85\textwidth]{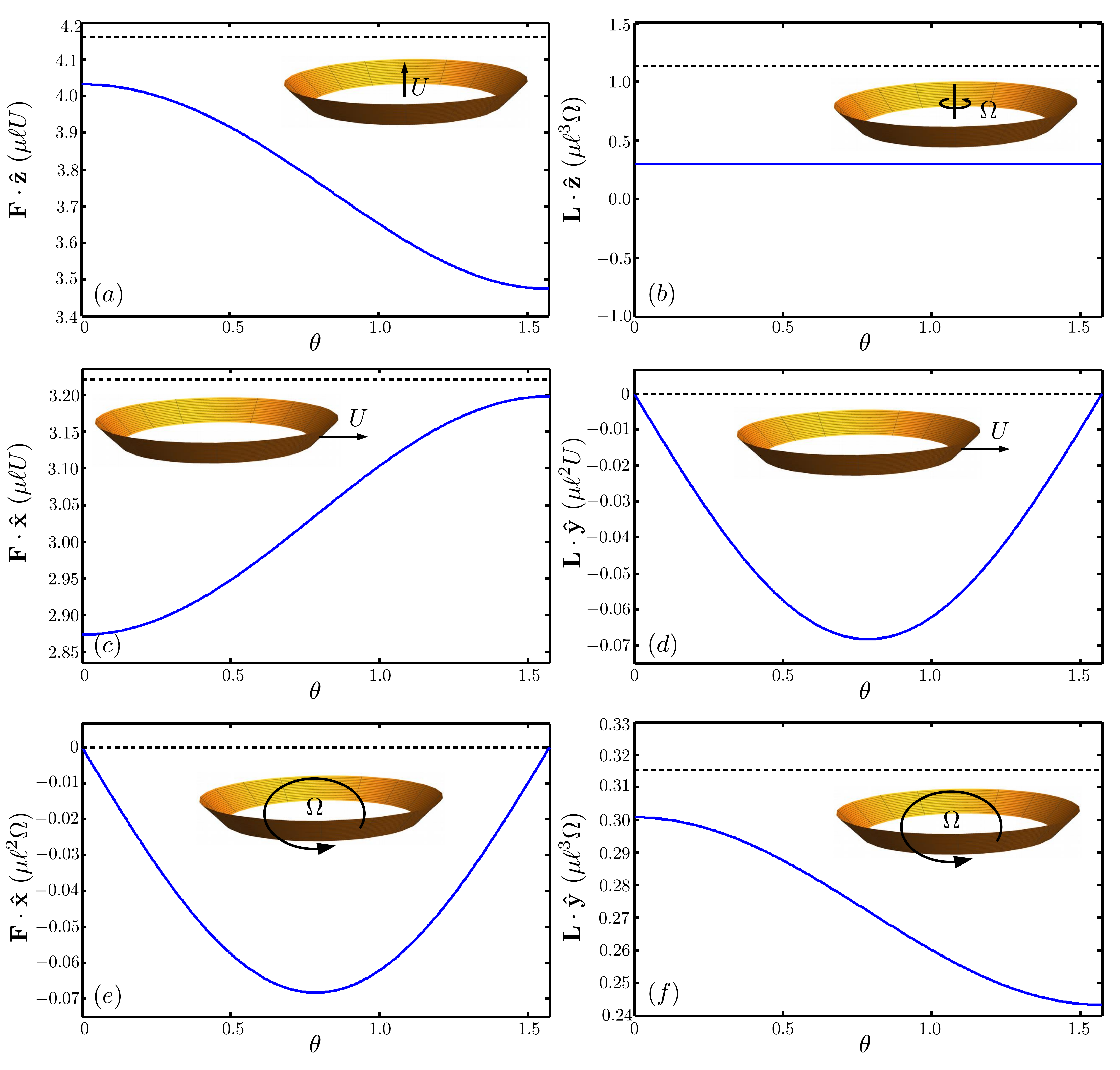}
\caption{The non-zero resistance coefficients of a ribbon torus (solid lines) and a torus filament (dashed lines) as a function of the angle $\theta$. (a) The force along $\mathbf{\hat{z}}$ from motion in the same direction; (b) The torque from rotation in $\mathbf{\hat{z}}$; similarly (c) and (d) plot the force and torque from translation in $\mathbf{\hat{x}}$, respectively; and (e) and (f) show the force and torque from rotation in $\mathbf{\hat{y}}$ respectively. Both the ribbon torus and the torus filament have  have $r_{b}/\ell =b_{\ell}=10^{-2}$. }
\label{fig:torusvalues}
\end{figure}

We first consider rigid translation in $\mathbf{\hat{z}}$, in which case the system is axisymmetric. Therefore the components of the force are constant, $f_{1,i}(s_{1}) = C_{i}$ and the equations become 
\begin{equation}
\left( \begin{array}{c}
0 \\
0 \\
8 \pi U
\end{array} \right) = \pi \left[ \boldsymbol{\Gamma}+ \boldsymbol{\Theta}  \right]\left( \begin{array}{c}
 C_{t} \\
 C_{n} \\
 C_{b}
\end{array} \right),
\end{equation}
where 
\begin{eqnarray}
 \notag \Gamma_{ij} &=& \int_{-1}^{1} \,d q \mbox{ } \mathbf{\hat{x}}_{i}(s_{1})\cdot   \left[ \frac{ \mathbf{I} + \mathbf{\hat{R}_{0}} \mathbf{\hat{R}_{0}}}{|R_{0}|} \cdot \mathbf{\hat{x}}_{j}(s_{1}+q)   -  \frac{ \left(\mathbf{I} + \mathbf{\hat{t}} \mathbf{\hat{t}}\right)}{|q|} \cdot  \mathbf{\hat{x}}_{j}(s_{1})  \right]  \\
 &\equiv & \left( \begin{array}{c c c}
4 L-6&0 &0 \\
0&2 L-6&0 \\
0&0&2 L
\end{array}\right), \\
 \boldsymbol{\Theta} &=& 
L_{2} \left(\mathbf{I}+\mathbf{\hat{t}}\mathbf{\hat{t}}\right) +2 \left(\mathbf{\hat{T}}\mathbf{\hat{T}} - \mathbf{\hat{t}}\mathbf{\hat{t}}\right),
\end{eqnarray}
and $\mathbf{\hat{x}}_{i}(s_{1})$ represents $\mathbf{\hat{t}}$, $\mathbf{\hat{n}}$ or $\mathbf{\hat{b}}$.
Solving this equation the total force and torque on the torus is\begin{eqnarray}
\mathbf{F}_{h} &=&  \frac{8 \pi U \left[\cos (2 \theta )+2L_{3}-5 \right]}{ 3 \cos (2 \theta )-2 L_{3}\left(2-L_{3}\right)-3 }  \mathbf{\hat{z}}, \label{torusz} \\
\mathbf{L}_{h} &=& \mathbf{0},
\end{eqnarray}
where
\begin{equation}
\beta_{1}= 6 \cos (2 \theta )+4 L^2+4 L (L_{2}-2)+(L_{2}-4) L_{2}-6,
\end{equation}
 $L=\ln(4/\pi)$, $L_{2} =\ln(16/b_{\ell}^{2})$, and $L_{3} =\ln({16}/{\pi b_{\ell}})$. The above result shows that the force from translation in $\mathbf{\hat{z}}$ 
 is maximised when $\theta=0$ ($\mathbf{\hat{T}} = \mathbf{\hat{n}}$) and minimized when $\theta=\pi/2$ ($\mathbf{\hat{T}} = \mathbf{\hat{b}}$) with a roughly sinusoidal dependence between the two. This is shown clearly in Fig.~\ref{fig:torusvalues}a where we plot all resistance coefficients of the ribbon torus in the case $b_{\ell}=10^{-2}$ and compared it to that of a slender filament.

\subsubsection{Rotation around torus axis ($\mathbf{\hat{z}}$)}

We now turn to the other axisymmetric motion of a torus, rotation in $\mathbf{\hat{z}}$.
For rotation around $\mathbf{\hat{z}}$ the force distribution is still constant but the surface velocity is now $\mathbf{U}=\Omega/\pi  \mathbf{\hat{t}}$.  
The equations to solve are
\begin{equation}
\left( \begin{array}{c}
8 \Omega \\
0 \\
0
\end{array} \right) =\pi \left[ \boldsymbol{\Gamma}+ \boldsymbol{\Theta} \right]\left( \begin{array}{c}
 C_{t} \\
 C_{n} \\
 C_{b}
\end{array} \right),
\end{equation}
which gives a net force and torque of
\begin{eqnarray}
\mathbf{F}_{h} &=& \mathbf{0}, \\
\mathbf{L}_{h} &=&   \frac{4 \Omega }{\pi \left(L_{3} -2\right)} \mathbf{\hat{z}} . 
\end{eqnarray} 
Therefore, to leading order, the resistance felt from the rotation around $\mathbf{\hat{z}}$ is independent of the slant of the ribbon, $\theta$ (Fig.~\ref{fig:torusvalues}b). This is unsurprising since, regardless of how $\mathbf{\hat{T}}$ is orientated, the surface is always moving along the tangent direction.

\subsubsection{Translation perpendicular to torus axis ($\mathbf{\hat{x}}$)}

We next consider linear translation in $\mathbf{\hat{x}}$, while the result for translation along $\mathbf{\hat{y}}$  can be deduced through a rotation of $\pi/2$. 
When a torus moves in $\mathbf{\hat{x}}$ the system is no longer axisymmetric and the velocity is given by
\begin{equation}
 U \mathbf{\hat{x}} = - U \sin(\pi s_{1}) \mathbf{\hat{t}} - U \cos(\pi s_{1}) \mathbf{\hat{n}}.
\end{equation}
The sinusoidal nature of this velocity suggests that the force coefficients should take the form
\begin{equation}
f_{1,i}(s_{1}) = C_{c,i}  \cos(\pi s_{1})  +C_{s,i} \sin(\pi s_{1}). \label{forcedisp}
\end{equation}
The orthogonality of the trigonometric functions then reduces the SRT equations to
\begin{equation}
-8 U \left( \begin{array}{c}
0 \\
1 \\
0 \\
1\\
0 \\
0
\end{array} \right) = 
\left( \begin{array}{c c}
\boldsymbol{\Gamma}^{cc} + \boldsymbol{\Theta}  & \boldsymbol{\Gamma}^{cs} \\
 \boldsymbol{\Gamma}^{sc} &  \boldsymbol{\Gamma}^{ss} +\boldsymbol{\Theta} 
\end{array}\right) 
\left( \begin{array}{c}
C_{c,t} \\
C_{c,n} \\
C_{c,b} \\
C_{s,t} \\
C_{s,n} \\
C_{s,b}
\end{array} \right) \label{perptorus},
\end{equation}
where
\begin{eqnarray}
 \boldsymbol{\Gamma}^{cc} = \boldsymbol{\Gamma}^{ss} &=&  \left( \begin{array}{c c c}
4 L -6 & 0&0  \\
0 & -2(L+1) & 0\\
0 &0 & 2(L-2)
\end{array}\right) , \\
\boldsymbol{\Gamma}^{cs} &=&  \left( \begin{array}{c c c}
0& 4 &0 \\
4 & 0 & 0\\
0 &0 & 0
\end{array}\right) ,
\\
 \boldsymbol{\Gamma}^{sc} &=& \left( \begin{array}{c c c}
0& 4 &0 \\
-4 & 0 & 0\\
0 &0 & 0
\end{array}\right),
\end{eqnarray}
and these $\boldsymbol{\Gamma}$ tensors are derived from the outer integral using the orthogonality of trigonometric functions; i.e. the value of $\boldsymbol{\Gamma}^{sc}$ is found by multiplying the outer integral by $\sin(\pi s_{1})$ and integrating over $s_{1}$, when only the $\cos(\pi s_{1})$ component of the force was considered. The other $\boldsymbol{\Gamma}$ tensors above are found similarly with the first letter in their superscript representing the multiplying trigonometric function, $s \equiv \sin(\pi s_{1})$ and $c\equiv\cos(\pi s_{1})$, and the second superscript representing the considered component of the force, $s \equiv \sin(\pi s_{1})$ and $c\equiv\cos(\pi s_{1})$.  
Solving the above equation the total force and torque on the translating ribbon torus is
\begin{eqnarray}
\mathbf{F}_{h} &=& -\frac{8 \pi  [6 L_{3}^{2} - 2L_{3}(2 L+13) + 6L- (L_{2}-7) \cos (2 \theta )+25]}{\beta_{2}} U \mathbf{\hat{x}} \label{torusx},\\
\mathbf{L}_{h} &=&\frac{16   ( L_{3}-3) \sin (2 \theta )}{\beta_{2}}  U \mathbf{\hat{y}} \label{torusxc},
\end{eqnarray} 
where 
\begin{eqnarray}
\notag \beta_{2} &=& 4 \left\{ \left[ L_{3}\left(1-2 L\right)+4 L-4\right]\cos (2 \theta )  -2 L_{3}^3+4 L_{3}^2 (2+L) \right.\\
 &&\left. - L_{3} (14 L +5)+ 4(3 L -1) \right\}.
\end{eqnarray}

A slender ribbon moving in the plane therefore experiences  a net force in the direction of motion and a torque perpendicular to the motion (still in the plane). The magnitude of the force felt is   minimized for $\theta=0$ and maximised for $\theta=\pi/2$ converse to the behaviour seen for translation in $\mathbf{\hat{z}}$ (see Fig.~\ref{fig:torusvalues}c).
  The non-zero torque depends on the orientation of the ribbon through $\sin(2\theta)$ and so is zero when the ribbon width is aligned with $\mathbf{\hat{n}}$ or $\mathbf{\hat{b}}$ (as expected by symmetry), and is maximised when $\theta=\pi/4$ (Fig.~\ref{fig:torusvalues}d). This torque is due to the asymmetric displacement of the fluid over the ribbon when it is slanted. 

\subsubsection{Rotation perpendicular to torus axis ($\mathbf{\hat{y}}$)}

The final motion to consider is rotation around $\mathbf{\hat{y}}$ (here also, rotation around $\mathbf{\hat{x}}$ may be deduced by symmetry). 
The velocity for rotation around $\mathbf{\hat{y}}$ is given by 
\begin{equation}
\mathbf{U} = - \Omega \frac{\cos(\pi s_{1}) }{\pi } \mathbf{\hat{b}}, \label{vely}
\end{equation}
and the force is again decomposed as in Eq.~\eqref{forcedisp}. 
The slender-ribbon equations therefore become
\begin{equation}
- \frac{\Omega}{\pi} \left( \begin{array}{c}
0 \\
0 \\
1 \\
0\\
0 \\
0
\end{array} \right) = 
\left( \begin{array}{c c}
\boldsymbol{\Gamma}^{cc} + \boldsymbol{\Theta}  & \boldsymbol{\Gamma}^{cs} \\
 \boldsymbol{\Gamma}^{sc} &  \boldsymbol{\Gamma}^{ss} +\boldsymbol{\Theta} 
\end{array}\right) 
\left( \begin{array}{c}
C_{c,t} \\
C_{c,n} \\
C_{c,b} \\
C_{s,t} \\
C_{s,n} \\
C_{s,b}
\end{array} \right), \label{perprottorus}
\end{equation}
Hence the net force and torque on the torus is
\begin{eqnarray}
\mathbf{F}_{h} &=& \frac{16    ( L_{3}-3)\sin (2 \theta )}{\beta_{2}} \Omega \mathbf{\hat{x}} ,\\
\mathbf{L}_{h} &=&   -\frac{16 \left[ (L_{3}-2) \cos (2 \theta ) +2 L_{3}^{2} -  L_{3}(4L+5) + 2(4L-1) \right]}{ \pi \beta_{2}}\mathbf{\hat{y}} \Omega \label{torusy}.
\end{eqnarray} 

These results are illustrated numerically in Fig.~\ref{fig:torusvalues}e and f. We see the same coupling between the force and rotation in $\mathbf{\hat{y}}$ as for torque and motion in $\mathbf{\hat{x}}$ as expected from the symmetries of the resistance matrix. Furthermore, the results reveal  a sinusoidal dependence on torque, which is maximal  at $\theta=0$ and minimal at $\theta=\pi/2$.

\subsubsection{Comparison to a slender torus}

The ribbon torus is the ribbon extension to a cylindrical torus. These shapes have been thoroughly studied and have a resistance matrix, to order $r_{b}/\ell$, of \cite{Johnson1979a}
\begin{equation}
\mathbf{R}_{slender} =  \left(\begin{array} {c c c c c c}
 \frac{2 \pi (6 L_{4} -17 )}{(2 L_{4}-1)(L_{4}-2) -4 } &0 &0 & 0 &0 &0\\
0&\frac{2 \pi (6 L_{4} -17 )}{(2 L_{4}-1)(L_{4}-2) -4 }  & 0 & 0 &0 &0 \\
0& 0&  \frac{16 \pi }{ 2 L_{4}+1 }& 0 &0 &0\\
0&0&0& \frac{8}{\pi (2 L_{4} -3)} &0 &0 \\
0&0&0&0&\frac{8}{\pi (2 L_{4} -3)} & 0 \\
0&0&0&0& 0&  \frac{4  }{L_{4} -2}
\end{array}  \right),
\end{equation}
where $L_{4} =  \ln\left( {8 \ell}/{\pi r_{b}}\right)$ and $r_{b}$ is the radius of the cylinder. We plot in Fig.~\ref{fig:torusvalues}  the cylindrical torus coefficients (dashed lines) for $r_{b}/\ell=b_{\ell}=10^{-2}$. These coefficients are simpler than that of the ribbon torus but show a similar logarithmic dependence on body's aspect ratio. Furthermore, these resistance coefficients are seen to be systematically larger that of a slender ribbon with the same ratio $b_{\ell}=r_{b}/\ell$ (Fig.~\ref{fig:torusvalues}). This is expected as the surface area of a cylinder with radius $r_{b}$ is greater  than that of a thin ribbon with width $r_{b} = b_{\ell} \ell$; hence the cylinder should experience greater drag. 

\subsubsection{The sedimentation of a ribbon torus}

\begin{figure}[t]
\centering
\includegraphics[width=.9\textwidth]{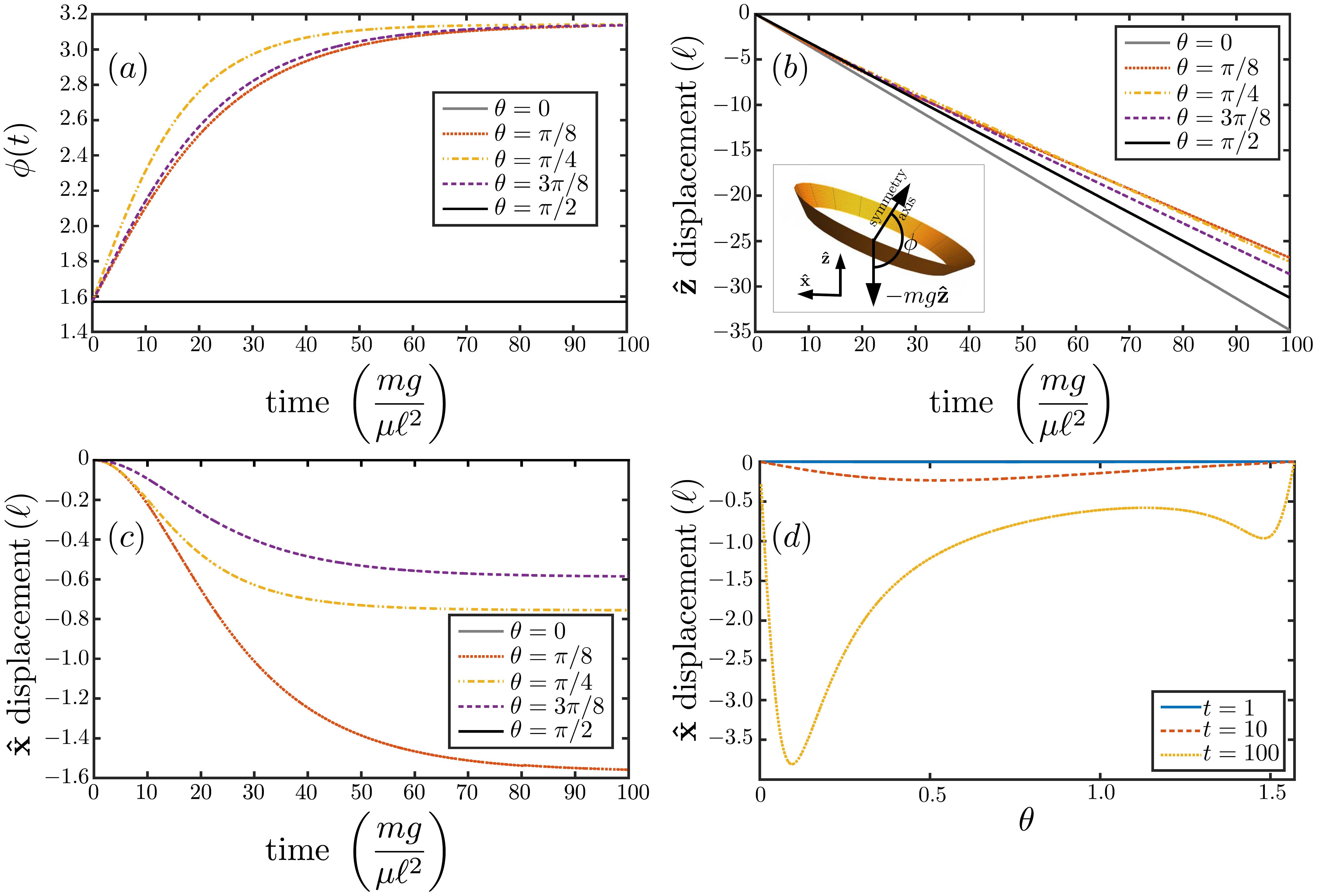}
\caption{Settling of a ribbon torus for different $\theta$. (a) The angle between the torus axis of symmetry and gravity; (b) The position in $\mathbf{\hat{z}}$ divided by time; (c) The position in $\mathbf{\hat{x}}$;  (d) The displacement experienced for different $\theta$ at $t=1$, 10 (dashed), and 100 (dotted). In (a), (b) and (c) the gray solid line is $\theta=0$, the red dotted line is $\theta = \pi/8$, the yellow dash-dotted line is $\theta=\pi/4$, the purple dashed line is $\theta=3 \pi/8$, and the solid black line is $\theta=\pi/2$. All plots use $b_{\ell}=0.01$, $mg=1$, scaled time and start with $\phi(0)=\pi/2$, $\mathbf{X}(0)=\mathbf{0}$. The inset in (b) demonstrates the orientation of the settling torus.}
\label{fig:sedimenttorus}
\end{figure}

The resistance coefficients obtained above allow us to consider the settling dynamics of a ribbon torus. The motion of a settling ribbon torus is two dimensional but for non-trivial values of $\theta$ the translation-rotation coupling will rotate the body as it settles. We orientate the ribbon such that the angle between its axis of symmetry and the gravitational force, $-mg\mathbf{\hat{z}}$,  is given by $\phi$ (Fig.~\ref{fig:sedimenttorus}b inset). Keeping the motion in the $\mathbf{\hat{z}}$-$\mathbf{\hat{x}}$ plane the velocity of the torus in the laboratory frame is given by
\begin{eqnarray}
\mathbf{U}_{sed} &=& \frac{mg}{R_{a,z} (R_{b}^{2} - R_{c} R_{a,x})}  \left(\begin{array}{c}
\left[R_{b}^{2} +(R_{a,z}-R_{a,x})R_{c} \right] \cos(\phi) \sin(\phi)\\
 0\\
R_{a,z} R_{c} \sin^{2}(\phi) -(R_{b}^{2} - R_{c} R_{a,x})\cos^{2}(\phi) 
\end{array} \right),\\
\boldsymbol{\Omega}_{sed} &=&  -\frac{mg }{R_{b}^{2} - R_{c} R_{a,x}}\left(\begin{array}{c}
 0\\
R_{b} \sin(\phi)\\
 0
\end{array} \right),
\end{eqnarray}
for a fixed value of $\phi$. In the above equations $R_{a,z}$ is the resistance coefficient relating force and translation parallel to the axis of symmetry, Eq.~\eqref{torusz}, $R_{a,x}$ is the coefficient relating force and translation perpendicular to the axis of symmetry, Eq.~\eqref{torusx}, $R_{b}$ is the coupling resistance coefficient, Eq.~\eqref{torusxc}, and $R_{c}$ is the rotational resistance coefficient, Eq.~\eqref{torusy}.

Given the motion is two dimensional we can then write the evolution equation for $\phi$ as
\begin{equation}
\frac{d \phi}{d t} = -\frac{mg R_{b} \sin(\phi) }{R_{b}^{2} - R_{c} R_{a,x}},
\end{equation}
which has the solution
\begin{equation}
\phi(t) = 2 \mbox{ acot}\left[\exp\left(\frac{mg R_{b}  t}{R_{b}^{2} - R_{c} R_{a,x}} \right) \right],
\end{equation}
where $\mbox{acot}$ represents the inverse of the cotangent and we have set $\phi(0) =\pi/2$. This function asymptotes to either $\phi =0$ for $R_{b}/(R_{b}^{2} - R_{c} R_{a,x})>0$, or $\phi=\pi$ for $R_{b}/(R_{b}^{2} - R_{c} R_{a,x})<0$, in the limit $t\rightarrow \infty$. Therefore for non-trivial  values of $\theta$ the ribbon torus will rotate to make its axis of symmetry parallel to the gravitational force. 

Inserting $\phi(t)$ into $\mathbf{U}_{sed}$ and integrating, the change in position is obtained analytically for all times as
\begin{equation}
\mathbf{X}(t) -\mathbf{X}(0) = \frac{1}{R_{a,z} R_{b}}\left(\begin{array}{c}
 \left[R_{b}^{2} +(R_{a,z}-R_{a,x})R_{c}\right] \left[\mbox{sech}\left(\frac{mg R_{b}  t}{R_{b}^{2} - R_{c} R_{a,x}}\right) -1\right]\\
 0\\
 - mg R_{b}  t + \left[R_{b}^{2} +(R_{a,z}-R_{a,x})R_{c}\right]\tanh\left(\frac{mg R_{b}  t}{R_{b}^{2} - R_{c} R_{a,x}}\right)
\end{array} \right),
\end{equation}
where $\mbox{sech}(x) = 1/\cosh(x)$. This position equation inherently assumes that $R_{b} \neq 0$. If $R_{b} =0$ the system would behave similar to a settling rod and so would be trivial. At relatively long times the torus translates in the direction of gravity while for shorter times it translates in both $\mathbf{\hat{x}}$ and $\mathbf{\hat{z}}$. In Figs.~\ref{fig:sedimenttorus}a, b, and c we show the change in orientation and position with time for different $\theta$ and in the case $b_{\ell}=0.01$. In particular, Fig.~\ref{fig:sedimenttorus}b shows that the displacement in $\mathbf{\hat{z}}$ initially starts off similarly to a torus falling in along its side but then slows down as the axis of symmetry aligns with gravity. This is a result of the  drag for motion parallel to the axis of symmetry being larger then the drag for motion perpendicular. 

Similarly the displacement in $\mathbf{\hat{x}}$ is seen to only occur during the reorientation, as would be expected. In addition, the $\mathbf{\hat{x}}$ displacement is also seen to have a maximum between $\theta=0$ and $\theta=\pi/4$ (Fig.~\ref{fig:sedimenttorus}d). This is due to tori for which $\theta$ close to $0$ or $\pi/2$ rotating slower, thereby giving longer displacement times, and the displacement rate in $\mathbf{\hat{x}}$ going to zero as $\phi \rightarrow \pi/2$. We note that there is a second smaller peak between $\theta=\pi/4$ and $\theta=\pi/2$. This is caused by the same features; however the drag from translation perpendicular to the axis of symmetry is higher for $\theta$ between $\pi/4$ and $\pi/2$ thereby reducing the net displacement.

\section{Conclusion} \label{conclusion}
Slender-ribbon theory provided a means to investigate the hydrodynamics of a wide class of ribbon configurations numerically \cite{Koens2016}. In this paper we showed that  the force distribution across the width of an  isolated ribbon located in a infinite fluid  can be determined analytically, irrespective of how the ribbon twists and turns. This reduces the surface integrals in the slender-ribbon theory equations to a line integral which is commonly calculated to determine the hydrodynamics of slender filaments (Eqs.~\ref{SRTred} and \ref{SRTring}). This reduction makes slender-ribbon theory much easier to implement. Note that  when other bodies, or surfaces, are present, hydrodynamic interactions  will change the force distribution across the ribbon's width (i.e.~in the  $s_{2}$ direction), potentially making the problem intractable analytically. 

 The reduction in complexity has then allowed analytical solutions to slender-ribbon theory to be found. This was done for a long flat ellipsoid and a ribbon torus. The resistance coefficients for a long flat ellipsoid matched the values reported in the literature \cite{Batchelor2006}, could be used to create a resistive-force theory for ribbons, and allowed us to characterise their sedimentation under gravity.   The force and torque on a ribbon torus, however, exhibited a sinusoidal dependence on the plane in which the ribbon sits and exhibited coupling between force and rotation. This coupling caused a settling ribbon torus to rotate as it settles,  aligning the axis of symmetry with the direction of gravity. 
 Furthermore, while the resistance coefficients of a ribbon torus are algebraically tedious when compared to the resistance coefficients for a slender cylindrical torus, they have been all derived analytically and show a similar logarithmic dependence on  body's aspect ratio.

 The simplification of the equations of slender-ribbon theory and the development of a ribbon resistive-force-theory  will allow the dynamics of various new physical and biophysical problems to be tackled.  For example, some eukaryotic microorganisms are known to use ribbon-like swimming appendages, called flagella vanes \cite{Leadbeater2015}. It is still unclear if these provide any fitness advantage to the cells, an issue which could be addressed using slender-ribbon theory. In the physical world,  the reduction   the equations of slender-ribbon theory to a single line integral will  allows the elasto-hydrodynamics of slender ribbons to be characterised  similarly to the classical problem of  elasto-hydrodynamics of slender filaments \cite{Wiggins1998}.

 \section*{Acknowledgements}
 This research was funded in part by the European Union through a Marie Curie CIG Grant (EL), an ERC Consolidator grant (EL) the Cambridge Trusts (LK), the Cambridge Philosophical society (LK) and the Cambridge hardship fund (LK). 

 \appendix
\section{Slender-ribbon theory for looped ribbons} \label{sec:a2}

The slender-ribbon theory derived in Ref.~\cite{Koens2016} determined the leading order hydrodynamics for finite bodies with ellipsoidal ends. If instead we wanted the hydrodynamics of a looped ribbon the resulting equations will be different. Much of the derivation is the same and so we will only point out the differences here. When the ribbon forms a closed loop $\mathbf{R}$ remains the same but $s'_{1}$ is replaced with $s_{1} + q$. The integrals over $s'_{1}$ are then replaced with integrals over $q$. The regions to expand in remain the same and exhibit no difference in the expanded kernels. The integrals within the different regions now take the form
\begin{equation} \label{A2}
\int_{-1}^{1} \frac{\chi^{i}}{\sqrt{\chi^{2} + h^{2} }^{j}} \,dq
\end{equation}
where $\epsilon \chi = q$, $h$ is an arbitrary function that does not depend on $q$, $i$ and $j$ are positive integers and $\epsilon$ is a small parameter. These integrals can be evaluated exactly and then expanded to get their asymptotic behaviour. To do so  we make the substitution $\chi = h \sinh(\phi)$ and reduce the equations to
\begin{equation}
\epsilon \int_{\mbox{arcsinh}(\frac{-1}{\epsilon h })}^{\mbox{arcsinh}(\frac{1}{\epsilon h})}  h^{i-j+1} \frac{ \sinh^{i}(\phi) }{ \cosh^{j-i}(\phi)} \,d\phi.
\end{equation}
These integrals have known solutions \cite{I.S.GradshteynAuthorI.M.RyzhikAuthorAlanJeffreyAuthor2000}  and table~\ref{tab:A2} lists the relevant leading order terms. 
\begin{table}
\centering
\begin{tabular}{l c c c }  \hline
 & i=0 & i=1 & i=2  \\ \hline
j=1 & $\epsilon \ln\left( \frac{4}{\epsilon^{2} h^{2} }\right) + O (\epsilon^{2})$ &   &     \\ 
j=2 & $\frac{\epsilon \pi}{h} +  O(\epsilon^{2})$ & $O(\epsilon^{2})$ &    \\ 
j=3 & $\frac{2 \epsilon }{h^{2}} + O(\epsilon^{2})$ & $ O(\epsilon^{3})$&$\epsilon \left[\ln\left( \frac{4}{\epsilon^{2} h^{2}}\right) -2 \right] +O (\epsilon^{3})$   \\ \hline
\end{tabular}
\caption[Table of asymptotic integral forms needed for looped slender ribbons]{Table of asymptotic integral forms of Eq.~\eqref{A2} for the looped ribbon SRT expansion.  }
\label{tab:A2}
\end{table}
Using these asymptotic integrals, the equation to describe the leading order hydrodynamics of a looped ribbon becomes
\begin{eqnarray}  
8 \pi \mathbf{U }(s_{1},s_{2}) &=& \int_{-1}^{1} \,d q  \left[ \frac{ \mathbf{I} + \mathbf{\hat{R}_{0}} \mathbf{\hat{R}_{0}}}{|R_{0}|} \cdot \rho(s_{1}+q) \left\langle\mathbf{f}\right\rangle(s_{1}+q)   -  \frac{ \left(\mathbf{I} + \mathbf{\hat{t}} \mathbf{\hat{t}}\right)}{|q|} \cdot  \rho(s_{1}) \left\langle\mathbf{f}\right\rangle(s_{1})  \right] \notag \\
 &&+  \int_{-1}^{1} \,d s'_{2} \ln\left(\frac{4}{b_{l}^{2} \rho^{2}(s_{1}) (s_{2}-s'_{2})^{2}}\right)  \left(\mathbf{I} +\mathbf{\hat{t}} \mathbf{\hat{t}} \right) \cdot  \rho(s_{1}) \mathbf{f}(s_{1},t_{2})\notag\\ 
 &&+2  \left(\mathbf{\hat{T}} \mathbf{\hat{T}}- \mathbf{\hat{t}} \mathbf{\hat{t}}\right) \cdot  \rho(s_{1}) \left\langle\mathbf{f}\right\rangle(s_{1}) 
 + O(b_{l}) + O(a_{l}),
\end{eqnarray}
where $\rho> 0$ for all $s_{1}$. These equations differ to the finite SRT equations in three ways: (a) all locations with $s'_{1}$ has been replaced with $s_{1}+q$; (b) the integrals over $s'_{1}$ are now over $q$; and (c) the logarithm no longer has a $1-s_{1}^{2}$ term within it. Note that these loops can take any form desired, provided the curvature does not become too large. 
 
\bibliographystyle{unsrt}
\bibliography{library}

\begin{thebibliography}{10}
\bibitem{Errors}
The form of these corrections terms were not explicitly stated in the original
  paper but they can be derived trivially using Taylor series. Furthermore when
  relevant to the expansion's applicability these error terms were discussed in
  detail.
\end{thebibliography}
\end{document}